\documentclass[aps,pra,twocolumn,nofootinbib,groupedaddress,floatfix]{revtex4-2}
\usepackage{graphicx}
\usepackage{amssymb}
\usepackage{amsmath}
\usepackage{amsfonts}
\usepackage{multirow}
\usepackage{longtable}
\usepackage{hyperref}
\usepackage[version=3]{mhchem} 

\usepackage{siunitx}
\usepackage{dcolumn}
\usepackage{comment}
\newcolumntype{d}[1]{D{.}{\cdot}{#1}}
\newcolumntype{.}{D{.}{.}{-1}}
\newcolumntype{,}{D{,}{,}{-1}}

\begin{document}

\title{
Enhancement of parity-violating energy difference of H$_2 X_2$ molecules by electronic excitation}

\author{Naoya Kuroda}
\affiliation{Department of Micro Engineering, Kyoto University, Kyoto 615-8540, Japan}
\author{Takumi Oho}
\affiliation{Department of Micro Engineering, Kyoto University, Kyoto 615-8540, Japan}
\author{Ayaki Sunaga}
\affiliation{Institute for Integrated Radiation and Nuclear Science, Kyoto University, Osaka 590-0494, Japan}
\author{Masato Senami}
\affiliation{Department of Micro Engineering, Kyoto University, Kyoto 615-8540, Japan}
\email{senami@me.kyoto-u.ac.jp}

\date{\today}

\begin{abstract}

The parity-violating energy difference (PVED)
between two enantiomers of a chiral molecule
is caused by the weak interaction.
Because of the smallness of the PVED,
nonzero PVED is yet to be discovered in experimental searches.
To detect the PVED, the search for molecules with large PVED values is important.
Previously, one of the authors proposed that
the PVED may be significantly enhanced in ionized or excited states.
The significant enhancement of the PVED in some electronic excited states 
is proven in this study using \ce{H2$X$2} ($X$=O, S, Se, Te) molecules as examples.
The maximum enhancement was an about 360-fold increase for \ce{H2Se2}.
For the PVED calculation,
we employ the finite-field perturbation theory (FFPT) within the equation-of-motion coupled-cluster theory
based on the exact two-component molecular-mean field Hamiltonian.
The relation between the enhancement of the PVED and the contribution to the PVED from the highest occupied molecular orbital is also examined.
The effects of computational elements,
such as parameters related to the electron correlation and FFPT
on PVED values in excited states of H$_2 X_2$ molecules are studied.

\end{abstract}

\maketitle
\section{Introduction}

Parity symmetry is one of the most important concepts in fundamental physics.
Some particles are classified by the property under the parity transformation.
The pion, for example, has negative parity \cite{PDG}.
The violation of the product of parity and charge conjugation
is known to be essential for baryogenesis,
which is the process to produce the dominance of matter (baryon) over antimatter (antibaryon)
in our universe \cite{baryogenesis}.
Only the weak interaction,
among the four fundamental forces
(the strong and weak forces, the electromagnetic force, and the gravitation),
violates the parity symmetry.
For example, the $W$ gauge bosons, $W^{\pm}$, interact with the left-handed electron,
but not with the right-handed electron.
The weak interaction predicts the parity-violating energy difference (PVED)
between two enantiomers of a chiral molecule.
The possible link between the PVED and the homochirality on the earth 
is discussed frequently \cite{text:homochirality,homochirality,Bast:2011}.

This PVED is demonstrated to be very small for various enantiomers.
Quantum chemistry computations can predict this energy
and the values are in the range of $ 10^{-18}$-$10^{-12}$~eV, for 
H$_2 X_2$ ($X$=O, S, Se, Te) molecules 
\cite{Bast:2011,Senami:2019,Inada:2018,Shee:2016,Thyssen:2000,Laerdahl:1999}
and 
amino acid molecules \cite{Senami:2020,MacDermott:2009}.
Many experimental challenges have been proposed
for observations of the PVED of chiral molecules.
These proposals are related to
vibrational-rotational spectroscopy \cite{Letokhov}, NMR \cite{NMR}, and so on.
However, no PVED signature has been discovered so far.
Measurement of the vibrational frequency difference 
between two enantiomers of the CHFClBr molecule \cite{CHFClBr}
yields the most rigorous upper limit.
In this experiment, 
the measured target is not the PVED in the ground state but the vibrational frequency difference,
and the vibrational frequency difference
is suggested to relate the PVED of electronic energy 
as 
$E_{\rm el}^{\rm PV} / E_{\rm el} 
\sim E_{\rm vib}^{\rm PV} / E_{\rm vib} $~\cite{Letokhov}.

To capture the signature of the PVED,
it is important to find molecules with larger PVED.
In the paper \cite{Senami:2019}, 
ionization or electronic excitation of chiral molecules
was predicted to enhance the PVED by one order of magnitude or more.
This enhancement is due to the breaking of the cancellation between contributions 
to the PVED from each orbital.
Contributions to the PVED from each orbital in chiral molecules are reported to be canceled out each other \cite{Senami:2019, Laerdahl:1999, Schwerdtfeger:2005}, and this was a disappointing feature.
However, this cancellation can be a hint to find molecules with the large PVED.
The matrix elements of PVED contribution in some orbitals near the highest occupied molecular orbital (HOMO)
are larger than the sum of all orbitals.
Therefore, 
ionization or electronic excitation of chiral molecules
may disrupt the cancellation
and remarkably enhance the PVED.
In the work \cite{Senami:2019},
this prediction was checked for 
the doubly ionized state of the H$_2$Te$_2$ molecule.
However, the enhancement was only 10\% for this ionization.

In the present work,
the speculation of the drastic enhancement of the PVED by electronic excitation
is confirmed for H$_2 X_2$ ($X$=O, S, Se, Te) molecules by
high accuracy quantum chemical computation of electronic excitation.
Excited states were calculated by Equation-of-Motion Coupled-Cluster (EOM-CC) theory
based on the eXact 2-Component Molecular-Mean Field (X2Cmmf) Hamiltonian.
The enhancement is confirmed by carefully investigating the dependence on the computational method.


\ce{H2$X$2}-series molecules are
one of the often-used target molecules for the investigation and the test computation of new methodologies,
because of their simple structure among chiral molecules.
The calculations of the electronic structure of \ce{H2$X$2} were conducted at various levels of theory: 
the first report was conducted at the non-relativistic (NR) level by treating the spin-orbit interaction perturbedly \cite{Hegstrom:1980}, and followed by: Dirac-Hartree-Fock \cite{Laerdahl:1999}, one-component (1c) \cite{Berger:2005} and four-component (4c) DFT \cite{Bast:2011}, and 1c- \cite{Horny:2015} and 4c-correlation methods \cite{Thyssen:2000,Stralen:2005,Shee:2018}. 
These 4c calculations are based on Dirac-Coulomb Hamiltonian, 
and the contribution from the Breit term is also estimated at the 1e-ZORA level within the Breit-Pauli framework \cite{Berger:2008}. 
At the MP2 \cite{Stralen:2005} and CCSD(T) \cite{Thyssen:2000} levels,
calibration studies on the PVED in H$_2X_2$ molecules are also reported.
The calculation of core properties like the PVED would be sensitive to the correlation procedure, especially, the correlation of the core electrons. 
All calculations above were performed at the electronic ground state of neutral systems.

Quack proposed 
an experiment in which electronic excited states
were employed \cite{Quack:1986,Quack:1989}
(see also reviews \cite{Quack:1994,Quack:2002}). 
In the work, the oscillation between parity eigenstates by the PVED is used for the observation.
The initial parity eigenstate of a chiral molecule is created by excitation to and deexcitation from
an electronic excited state with an achiral geometry of the molecule.
In the course of this proposal,
Berger found that the excited state of formyl fluoride (CHFO) molecule 
has three times larger PVED than that of the ground state
in the same structure of the excited and ground state \cite{Berger:2003}.
In the work, the PVED was calculated by the second-order perturbation theory
in the NR framework,
and the denominator of the parity violating potential has a factor, $E_0 - E_i$,
where $E_0$ denotes the energy of the reference state (not necessarily the ground state)
and $E_i$ denotes that of an excited state, $i$.
The enhancement of the PVED was explained as follows:
The energy of the ground state of a closed-shell chiral molecule is 
far apart from those of other states, that is, the denominator is large,
while the energy of an excited state may be close to those of other states.
Our enhancement mechanism is distinct in various points from this mechanism.
Our mechanism is motivated by the cancellation between contributions to the PVED from each orbital
and gives orders of magnitude enhancement.
Furthermore, we study it in relativistic theory.
Hence, we regard our mechanism to be distinct,
although it may explain 
how Berger's enhancement arises in relativistic theory.

This paper is organized as follows.
In the next section, 
the definition of the PVED is introduced briefly.
Then,
our computational method and details are described in Sec.~\ref{sec:comp}.
In Sec.~\ref{sec:results}, 
our results are presented.
The dependences of the PVED on some parameters of computations are studied
for the H$_2$O$_2$ molecule as an example.
Then, the enhancement of the PVED by electronic excitation
is demonstrated for H$_2 X_2$ molecules.
The last section is devoted to our conclusion.

\section{Theory}

The PVED dominantly arises from the exchange of $Z$ gauge bosons 
between electrons and nuclei.
The two electron contributions are reported to be subdominant~\cite{Sapirstein:2002}.
The $Z$ boson is a massive vector particle,
whose mass is about 91.1876(21) GeV/$c^2$ \cite{PDG}.
Due to this heaviness this interaction can be described
as a contact interaction.
The Hamiltonian density of this contact interaction is expressed as 
\begin{align}
{\cal H} (x)
=&
\frac{ G_F }{ 2\sqrt{2}  } 
\bar \psi_e (x) \gamma^\mu (g_V^e - g_A^e \gamma_5) \psi_e (x)
\nonumber \\
&\times
\bar \psi_{n} (x) \gamma_\mu (g_V^{n} - g_A^{n} \gamma_5) \psi_{n} (x),
\end{align}
where $\psi_e$ and $\psi_{n}$ are field operators of the electron and the nucleus, $n$,
$\bar \psi_i$ means the Dirac conjugate, $\bar \psi_i \equiv \psi_i^\dagger \gamma^0$,
$ \gamma_5 $ is defined with gamma matrices as
$\gamma_5 = \gamma^5 \equiv i \gamma^0 \gamma^1 \gamma^2 \gamma^3 $,
and $G_F$ is the Fermi coupling constant,
$G_F /(\hbar c )^3 = 1.1663787(6)\times 10^{-5}$ GeV$^{-2}$ \cite{PDG}.
The neutral current couplings of the electron are denoted by 
$g_V^e = -1 + 4 \sin^2 \theta_W $ and $g_A^e = -1 $,
where $\theta_W$ is the weak-mixing angle, $\sin^2 \theta_W = 0.23121(4) $ \cite{PDG}.
The couplings of nuclei are given by 
$g_V^{n} = Z^{n} (1 - 4 \sin^2 \theta _W) - N^{n}$,
and $g_A^{n} = Z^{n} - N^{n}$,
where $Z^{n} $ and $ N^{n}$ are the number of protons and neutrons 
in the nucleus, $n$. 
Since $(1 - 4 \sin^2 \theta _W)$ is 0.0752, the dominant contribution to $g_V^{n}$ is the second term, $N^n$.

The parity violating contribution is derived as 
a cross term of $V$ and $A$ parts,
that is, the product of $g_V^e $ and $ g_A^{n} \gamma_5 $ parts
or that of $ g_A^e \gamma_5 $ and $ g_V^{n} $ parts.
The contribution from the former one is studied as
the subject of the parity violation in NMR experiments.
This contribution is dependent on nuclear spin and
for our $X$ nuclei 
isotopes with large natural abundance are singlet.
Therefore, the former contribution is neglected in this paper.
The latter contribution to the PVED is expressed
as follows:
\begin{align}
{\cal H}_{\rm PV}^{n}(x) 
= 
- \frac{ G_F }{ 2\sqrt{2}  } g_A^e g_V^{n}
\bar \psi_e (x) \gamma^\mu \gamma_5 \psi_e (x) 
\bar \psi_{n} (x) \gamma_\mu \psi_{n} (x) .
\end{align}
This can be reduced to the scalar form in the NR nucleus limit,
where the $\mu=1-3$ components of $\bar \psi_{n} \gamma_\mu \psi_{n} $ are negligible,
\begin{align}
{\cal H}_{\rm PV}^{n} (x) 
=
- \frac{ G_F }{ 2\sqrt{2} } g_A^e g_V^{n}
\psi_e^\dagger (x) \gamma_5 \psi_e (x) 
\psi_{n}^\dagger (x) \psi_{n} (x) .
\end{align}

The parity-violating energy is defined as 
the integration of the expectation value of this Hamiltonian density,
\begin{align}
E_{\rm PV}
= \int d^3 x \langle \Psi | \sum_{n} {\cal H}^{n}_{\rm PV} (x) |\Psi\rangle  ,
\end{align}
where the ket, $|\Psi\rangle$, is a state vector.
The PVED,
which is the energy difference between enantiomeric pair molecules,
is defined as twice the parity-violating energy, 
\begin{align}
\Delta E_{\rm PV}
= 2 | E_{\rm PV} |.
\end{align}
From the viewpoint of this relation, 
$E_{\rm PV}$ is studied in the following.
In this work, the sign of the enhancement is not paid attention to in the following,
since the PVED has a positive definition.

The interaction length of the weak interaction and 
the radius of a nucleus are very short,
and $ E_{\rm PV} $ can clearly be divided into the contributions 
from each nucleus,
\begin{align}
E_{\rm PV}
=& - \frac{ G_F}{ 2 \sqrt{2} }  \sum_{n} g_A^e g_V^{n}
\nonumber \\
&\times 
\left( \int d^3 x \langle \Psi | 
\hat \psi_{e}^\dagger (x) \gamma_5 \hat \psi_{e} (x) 
\hat \psi_{n}^\dagger (x) \hat \psi_{n} (x) | \Psi\rangle 
\right)
\nonumber 
\\
=& \frac{ G_F}{ 2 \sqrt{2} }  \sum_{n} g_V^{n}
 M_{\rm PV}^{n} .
 \label{Eq:EPV}
\end{align}
Here,
$ M_{\rm PV}^{n} $ parametrizes the contribution from the nucleus, ${n}$.
The density distribution of the nucleus is strongly localized
and, hence,
the contribution of the electron chirality density close to the nucleus, 
$ \langle \Psi | \hat \psi_e^\dagger \gamma_5 \hat \psi_e | \Psi \rangle $,
is dominant for the parity-violating energy.



\section{Computational Detail}
\label{sec:comp}

\subsection{Expectation value}

We used the following three methods for the computation of the expectation value of $E_{\rm{PV}}$.
i) The analytical one-electron integration at the Hartree-Fock (HF) method.
ii) The Z-vector equation at the CCSD method \cite{Shee:2016} for the ground state.
iii) The Finite-Field Perturbation Theory (FFPT) \cite{Pople:1968, Pawlowski:2015, Norman:2018} 
at the CCSD and EOM-CCSD method \cite{Shee:2018} for the ground and excited states.
In the relativistic coupled cluster framework, the calculations of $E_\textrm{PV}$ \cite{Thyssen:2000}, a P-violating property \cite{Hao:2018}, and P- and T- violating properties \cite{Abe:2018, Denis:2019, Denis:2020, Haase:2021} using the FFPT method have been reported. 
To the best of our knowledge, 
this work is the first report of the property calculation in excited states at the relativistic EOM-CC level.

In the FFPT method, we calculate the expectation value of a perturbation operator by numerically differentiating the total energy.
We first define the total Hamiltonian as follows:
\begin{align}
\hat H = {{\hat H}_0} + \lambda \hat O,
\end{align}
where ${\hat H}_0$ is the unperturbed Hamiltonian, $\lambda$ is the perturbation parameter, and $\hat O$ is the target operator.
In this study, $\lambda$ and $\hat O$ correspond to $\frac{ G_F}{ 2 \sqrt{2} }g_V^{n}$ and
$ \int d^3 x \hat \psi_{e}^\dagger (x) \gamma_5 \hat \psi_{e} (x) 
\hat \psi_{n}^\dagger (x) \hat \psi_{n} (x) $, respectively. 
We consider only O, S, Se, and Te for the atom $n$.
From the Hellmann-Feynman theorem, the derivative of the energy can be expressed as follows:
\begin{align}
{\left. {\frac{{\partial E\left( \lambda  \right)}}{{\partial \lambda }}} \right|_{\lambda  = 0}} 
&= \left\langle {\Psi \left| {\frac{{\partial \hat H}}{{\partial \lambda }}} \right|\Psi } \right\rangle 
\nonumber 
\\
&= \left\langle {\Psi \left| {\hat O} \right|\Psi } \right\rangle, 
\end{align}
where $\Psi$ and $E(\lambda)$ are the wavefunction and the energy with respect to the total Hamiltonian $\hat H$, respectively.
Once $\langle {\Psi | {\hat O} |\Psi } \rangle = M_{\rm PV}^n$
is derived by this method,
$E_{\rm PV}$ can be calculated in Eq. (\ref{Eq:EPV}).

In this study, we approximated the numerical derivative as follows:
\begin{align}
\label{eqn:numdev}
{\left. {\frac{{\partial E\left( \lambda  \right)}}{{\partial \lambda }}} \right|_{\lambda  = 0}} \approx \frac{{E\left( \lambda  \right) - E\left( { - \lambda } \right)}}{{2\lambda }}.
\end{align}
In this expression, we neglected the contribution from the order beyond $O(\lambda^3)$. 
Since the result is roughly independent of $ \lambda $
(the dependence will be discussed later),
the value of $\lambda$ was taken to be a variable free from $\frac{ G_F}{ 2 \sqrt{2} }g_V^{n}$.
To compute $ M_{\rm PV}^n $ accurately,
the appropriate value of $\lambda $ was set in our calculation.
We note that other four sources of errors that the above FFPT procedure leads: i) We neglected the contribution of the hydrogen atoms. ii) $\Psi$ is not the solution of the Hamiltonian $\hat H_0$, because of the unrealistically large contribution of $\lambda \hat O$. iii) The results depend on the value of $\lambda$ in Eq. (\ref{eqn:numdev}). iv) In the numerical derivative, more strict threshold would be needed than the default setup. 
The error due to i) is negligible, because the PVED rapidly increases as $Z$ increases with the ratio scaling roughly as $Z^5$ \cite{Zel:1977, Harris:1978}
and $g_V^n$ is small for the hydrogen nucleus (proton). We elucidate the effect of the errors ii) and iii) by comparing the results at the FFPT and the Z-vector method in the ground state by using CCSD method. We elucidate the effect of iv) by changing the norm of the residual vectors of EOM-CC calculation.  
We choose the set of the values of $\lambda$ referring to the previous study using the FFPT method for the ground state \cite{Thyssen:2000}.

\subsection{Computational models and parameters}

\begin{figure}[t]
        \centering
          \includegraphics[width=0.5\linewidth]{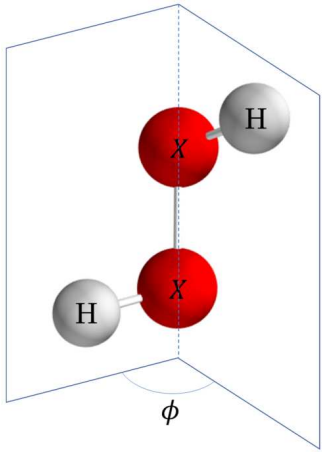} 
        \caption{Definition of the dihedral angle $\phi$.
       }
         \label{fig:H2X2_dihedral}
\end{figure}

We employed the molecular mean-field approximations \cite{Sikkema:2009} to the Dirac-Coulomb-Gaunt ($^2DCG_M$) Hamiltonian for the correlated calculations, and the Dirac-Coulomb-Gaunt at the HF levels.
For the calculation of the wavefunction, we used RELCCSD modules at the CCSD \cite{Visscher:1995, Visscher:1996} and EOM-CC \cite{Shee:2018} levels. (We give the information about the numbers of the correlated electrons and the active space in the next section.)
For these computations, we used DIRAC19 program package \cite{DIRAC19, DIRAC:2020}. 
We employed Dyall's relativistic basis sets of triple-zeta quality (dyall.ae3z) and augmented ones (dyall.aae3z, dyall.acv3z, and dyall.av3z) \cite{basis_set} in the uncontracted form for all atoms. 
The structure of the molecules was optimized at the HF method for the electronic ground state and dyall.ae2z basis sets at the Dirac-Coulomb Hamiltonian level by utilizing DIRAC code.
For excited states, the same structure were employed to derive all excited-state values in a single calculation.
For the optimization computations,
Visscher's approximation was used for the two-electron integration of (SS$\mid$SS) class \cite{Visscher:1997}.
In Table \ref{tab:H2X2_opt}, the optimized structures of ${\rm{H_2}}X_2$ molecules are summarized. 
The dihedral angle, $\phi$, is defined in Fig.~\ref{fig:H2X2_dihedral}.
The Gaussian charge distribution was employed as the nuclear model for both the nucleus-electron interaction and the nuclear charge density in the PV operator \cite{Visscher_Gauss:1997}.
We set the convergence criterion of the amplitudes of the CC calculations to $10^{-12}$, which is the default setting of DIRAC code. If not stated explicitly, the convergence threshold on the norm of the residual vectors at the EOM-CCSD stage was set to $10^{-10}$ for the numerical stability of the finite-field calculation.

\begin{table}[tbp]
\caption{Optimized structures of ${\rm{H_2}}X_2$ molecules. 
}
\centering
\scalebox{1.0}{
\begin{tabular}{ccccc}
\hline \hline
\begin{tabular}{c} ${\mathrm{H_2}} X_2$ \end{tabular} & 
\begin{tabular}{c} $X$-$X$ \\bond length \\{[}$\si{\angstrom}${]} \end{tabular} & 
\begin{tabular}{c} H-$X$ \\bond length \\{[}$\si{\angstrom}${]} \end{tabular} & 
\begin{tabular}{c} H-$X$-$X$ \\bond angle \\{[}deg{]} \end{tabular} & 
\begin{tabular}{c} Dihedral \\angle $\phi$\\{[}deg{]} \end{tabular} \tabularnewline
\hline 
${\rm {H_{2}O_{2}}}$ & $1.390$ & $0.944$ & $103$ & $115$\tabularnewline
${\rm {H_{2}S_{2}}}$ & $2.058$ & $1.332$ & $99$ & $90$\tabularnewline
${\rm {H_{2}Se_{2}}}$ & $2.333$ & $1.455$ & $97$ & $90$\tabularnewline
${\rm {H_{2}Te_{2}}}$ & $2.729$ & $1.650$ & $96$ & $90$\tabularnewline
\hline \hline
\end{tabular}
}
\label{tab:H2X2_opt}
\end{table} 

\section{Results}
In this section, we employ the atomic units, that is, Hartree ($E_h$) to express $E_{\rm PV}$.
\label{sec:results}

\subsection{Parameter dependence of $E_{\rm PV}$ in H$_2$O$_2$}
In this subsection, we investigate the dependence of $E_{\rm PV}$ on the perturbation parameter $\lambda$ in the FFPT approach by using the H$_2$O$_2$ molecule as an example. For more accurate calculations, we should obtain the most suitable $\lambda$ for each excited state and, however, this approach is expensive.
A practical approach is using the same $\lambda$ for all excited states of the same molecule at the EOM-CCSD level after estimating the error of this approach. 
We also investigate the dependence on the threshold, because the FFPT calculation is sensitive to numerical noise.
Effects of correlation and diffuse functions of basis sets on $E_{\rm PV}$ 
are studied by comparing several basis sets.

Table \ref{tab:lambda_compare_GS} demonstrates
the dependence of $E_{{\rm {PV}}}$ on $\lambda$ in the FFPT method at the electronic ground state.
In the Z-vector calculation, the contribution from the hydrogen was neglected,
since this contribution was also dropped in FFPT results.
From the table, it is found that the most suitable $\lambda$ is $1.0\times10^{-3}$ (a.u.)
with the difference from the Z-vector calculation, $-1.2\%$.
The differences for $\lambda =1.0\times10^{-2},1.0\times10^{-4}$ (a.u.) were similar in magnitude to that of $1.0\times10^{-3}$ (a.u.). 
When $\lambda$ was very small, $1.0\times10^{-5}$ (a.u.), the agreement with the reference value was bad (37.6\%).
If we employed $\lambda =1.0\times10^{-2},1.0\times10^{-4}$ (a.u.)
instead of the most suitable value $1.0\times10^{-3}$ (a.u.),
our conclusions about the estimate of the enhancement are unchanged.
Too large change beyond the range (that is, $1.0\times10^{-5}$ (a.u.)) may give rise to too large an error.
For other H$_2X_2$, suitable values of $\lambda$ were also determined by 
comparing with Z-vector results.
In the following, $\lambda = 10^{-3}, 10^{-4}$ and $10^{-5}$ (a.u.) are adopted
for H$_2$S$_2$, H$_2$Se$_2$, and H$_2$Te$_2$, respectively.
Errors of FFPT values obtained by utilizing these $\lambda$ are
considered to be less than 2\%
in the valence correlated calculations, as shown in the next section.

\begin{table}[tbp]
\caption{Dependence of $E_{{\rm {PV}}}$ on $\lambda$ within the FFPT method in the electronic ground state of $\rm{H_2 O_2}$. 
 Dev. represents the relative deviation of the FFPT result from the Z-vector one.
The dyall.aae3z basis set was used for all atoms.
All electrons were correlated and the virtual spinors were not truncated. 
}
\centering
\scalebox{1.0}{
\begin{tabular}{cccr}
\hline \hline 
\begin{tabular}{c} Method \end{tabular} & 
\begin{tabular}{c} $\lambda$ {[}a.u.{]} \end{tabular} & 
\begin{tabular}{c} $E_{{\rm {PV}}}/10^{-19}E_{h}$ \end{tabular} & 
\begin{tabular}{c} Dev. {[}\%{]} \end{tabular} \tabularnewline
\hline 
\multicolumn{1}{c}{FFPT} & $1.0\times10^{-2}$ & $4.240$ & $2.0$\tabularnewline
 & $1.0\times10^{-3}$ & $4.110$ & $-1.2$\tabularnewline
 & $1.0\times10^{-4}$ & $4.306$ & $3.6$\tabularnewline
 & $1.0\times10^{-5}$ & $2.595$ & $-37.6$\tabularnewline
\hline 
Z-vector & - & $4.158$ & - \tabularnewline
\hline \hline 
\end{tabular}
}
\label{tab:lambda_compare_GS}
\end{table}

Table \ref{tab:lambda_compare_ES} shows the dependence of $E_{{\rm {PV}}}$ on $\lambda$ in the FFPT approach for electronic excited states of $\rm{H_2 O_2}$. 
The mean values and the standard deviations of $E_{\rm PV}$ in ES1(a), ES1(b), ES2(a), and ES2(b) for $1.0\times10^{-n}$ (a.u.) ($n=2,3,4$) are $-1.5\pm0.7$, $14.3\pm2.0$, $1360.0\pm19.0$, and $-18.6\pm 1.2$ ($E_h$), respectively,
where ES$m(x)$ represents the $m$-th excited state obeying the $x$-symmetry. 
The standard deviation for $\lambda$ is much smaller than the enhancement of $E_{\rm PV}$,
except for ES1(a). 
The value of $\lambda$ suitable for the ground state computations 
was chosen for the following computations of excited states even if it might not be the best.
Our purpose is to establish the existence of the enhancement of $E_{\rm PV}$
by two or three orders of magnitude by electronic excitation.
Hence, potential error caused by $\lambda$ does not affect our conclusion.
For all $\lambda$ and even for $\lambda=1.0\times10^{-5}$ (a.u.),
$E_{{\rm {PV}}}$ in ES2(a) increased by about 300 times
compared to the ground state (GS).
We will discuss this enhancement later in detail.
For ES1(a), the standard deviation is similar order as that of the mean value.
However, our purpose is not the accurate computation, and
the accuracy of the state with small $E_{\rm PV}$ is not a serious issue.
The large difference in $E_{\rm PV}$ by $\lambda$ in the state with a small $E_{\rm PV}$
is speculated to be caused by numerical noise,
and this noise may be prevented by adopting a strict threshold value in some cases.

\begin{table}[tbp]
\caption{Dependence of $E_{{\rm {PV}}}$ on $\lambda$ within the FFPT method in the ground (GS),
first-excited (ES1), and second-excited (ES2) electronic states of $\rm{H_2 O_2}$. 
The dyall.aae3z basis set was used for all atoms.
All electrons were correlated and the virtual spinors were not truncated. 
}
\centering
\scalebox{1.0}{
\begin{tabular}{cccr}
\hline 
\hline 
\begin{tabular}{c}  $\lambda$ {[}a.u.{]} \end{tabular} & 
\begin{tabular}{c} State \end{tabular} &
\begin{tabular}{c} Symmetry \end{tabular} &
\begin{tabular}{c} $E_{{\rm {PV}}}/10^{-19}E_{h}$ \end{tabular}  \tabularnewline
\hline 
$1.0\times10^{-2}$ &GS               &a  &$4.240 $  \tabularnewline
                               &ES1      &a  &$-1.956 $  \tabularnewline
                               &                          &b&$12.168 $  \tabularnewline
                               & ES2 &a  &$1381.976 $  \tabularnewline
                               &                       &b&$-17.392 $  \tabularnewline
\hline 
$1.0\times10^{-3}$ &GS               &a  &$4.110 $  \tabularnewline
                               &ES1      &a  &$-1.902 $  \tabularnewline
                               &                         &b&$14.800 $  \tabularnewline
                               & ES2 &a  &$1348.683 $  \tabularnewline
                               &                       &b&$-19.839 $  \tabularnewline
\hline 
$1.0\times10^{-4}$ &GS             &a  &$4.306 $  \tabularnewline
                               &ES1      &a  &$-0.700 $  \tabularnewline
                               &                          &b&$16.018  $  \tabularnewline
                               & ES2 &a  &$1349.469 $  \tabularnewline
                               &                       &b&$-18.654  $  \tabularnewline
\hline 
$1.0\times10^{-5}$ &GS            &a  &$2.595 $  \tabularnewline
                               &ES1      &a  &$-34.637 $  \tabularnewline
                               &                          &b&$-17.931  $  \tabularnewline
                               & ES2 &a  &$1331.494  $  \tabularnewline
                               &                       &b&$-52.569 $  \tabularnewline
\hline 
\hline 
\end{tabular}
}
\label{tab:lambda_compare_ES}
\end{table}

Next, the dependence of $E_{{\rm {PV}}}$ on the threshold value in the EOM-CC calculation is checked,
and results are summarized in Table \ref{tab:threshold_compare}.
From the table, it is seen that the dependence is much weaker than that on $\lambda$ in Table \ref{tab:lambda_compare_ES} for all excited states.
For our purpose of this work, 
the dependence on the threshold is negligible.
Compared to $1.0\times10^{-12}$ (the best case),
large deviations were only $-$2.5\% and 2.0\% for ES1(b) and ES2(b) at the threshold $1.0\times10^{-9}$, respectively.
The value of the threshold $1.0\times10^{-10}$ is chosen for 
computations of $E_{\rm PV}$ in the following computations
from the viewpoint of the balance between the accuracy and computational cost, 
where the deviations from results of $1.0\times10^{-12}$ were less than 1\%. 
For other $X$ atoms, a larger threshold is speculated to be safely acceptable,
since larger spin-orbit interaction gives larger $E_{\rm PV}$.

\begin{table}[tbp]
\caption{Dependence of $E_{{\rm {PV}}}$ on the threshold value within the FFPT method in the ground (GS), first-excited (ES1), and second-excited (ES2) electronic states of $\rm{H_2 O_2}$.
	The dyall.aae3z basis set was used for all atoms.
	All electrons were correlated and the virtual spinors were not truncated. 
The perturbation parameter $\lambda$ was set to $10^{-3}$ (a.u.). 
}
\centering
\scalebox{1.0}{
\begin{tabular}{cccr}
\hline 
\hline 
Threshold& 
State  &
Symmetry &
$E_{{\rm {PV}}}/10^{-19}E_{h}$ \tabularnewline
\hline 
$1.0\times10^{-8}$  &ES1      &a  &$-1.822$  \tabularnewline
                               &                          &b&$14.882 $  \tabularnewline
                               & ES2 &a  &$1348.893$  \tabularnewline
                               &                       &b&$-19.762$  \tabularnewline
\hline 
$1.0\times10^{-9}$  &ES1      &a  &$-1.564$  \tabularnewline
                               &                         &b&$15.129 $  \tabularnewline
                               & ES2 &a  &$1348.887 $  \tabularnewline
                               &                       &b&$-19.488 $  \tabularnewline
\hline 
$1.0\times10^{-10}$  &ES1      &a  &$-1.902$  \tabularnewline
                               &                          &b&$14.800 $  \tabularnewline
                               & ES2 &a  &$1348.683 $  \tabularnewline
                               &                       &b&$-19.839 $  \tabularnewline
\hline 
$1.0\times10^{-11}$  &ES1      &a  &$-1.847$  \tabularnewline
                               &                          &b&$14.845 $  \tabularnewline
                               & ES2 &a  &$1348.806 $  \tabularnewline
                               &                       &b&$-19.771 $  \tabularnewline
\hline 
$1.0\times10^{-12}$  &ES1      &a  &$-1.941$  \tabularnewline
                               &                          &b&$14.760 $  \tabularnewline
                               & ES2 &a  &$1348.785 $  \tabularnewline
                               &                       &b&$-19.878 $  \tabularnewline
\hline 
\hline 
\end{tabular}
}
\label{tab:threshold_compare}
\end{table}

Finally, we study the contributions of the correlation and diffuse functions of the basis sets.
The significance of correlation and diffuse functions are shown in Table \ref{tab:basisset_compare}.
All basis sets in these calculations are the same quality, dyall triple-zeta basis sets,
and correlation and diffuse functions are different.
From the comparison between the values of dyall.av3z and dyall.aae3z,
large difference was found at ES2(a), $-$8.5\%.
Hence, correlation functions even for core electrons may affect the accuracy.
It is well known that diffuse functions play an important role in calculations of electronic excited states,
and it is significantly important in our study.
In the comparison between results in ES2(a) of dyall.ae3z and dyall.aae3z,
the value of the former was about three times that of the latter.
Hence, correlation and diffuse functions are important for our study.
For other $X$ atoms,
computations with the dyall.aae3z basis set are expensive.
Hence, the dyall.acv3z basis set is adopted for our computations
from the accuracy and computational cost perspective.
(For the oxygen atom, the dyall.acv3z basis set is the same as the dyall.aae3z basis set.)

\begin{table}[tbp]
\caption{Dependence of $E_{{\rm {PV}}}$ on basis sets within the FFPT method in the ground (GS),
first-excited (ES1), and second-excited (ES2) electronic states of $\rm{H_2 O_2}$.  
All electrons were correlated and the virtual spinors were not truncated.
The perturbation parameter $\lambda$ was set to $10^{-3}$ (a.u.). 
}
\centering
\scalebox{1.0}{
\begin{tabular}{cccr}
\hline 
\hline 
\begin{tabular}{c}Basis \end{tabular} & 
\begin{tabular}{c} State \end{tabular} &
\begin{tabular}{c} Symmetry \end{tabular} &
\begin{tabular}{c} $E_{{\rm {PV}}}/10^{-19}E_{h}$ \end{tabular}  \tabularnewline
\hline 
dyall.av3z              &GS              &a  &$4.096$  \tabularnewline
                               &ES1      &a  &$-1.840$  \tabularnewline
                               &                          &b&$15.054 $  \tabularnewline
                               & ES2 &a  &$1243.297 $  \tabularnewline
                               &                       &b&$-20.008 $  \tabularnewline
\hline 
dyall.ae3z               &GS               &a  &$4.162$  \tabularnewline
                               &ES1     &a  &$-2.546$  \tabularnewline
                               &                          &b&$16.365 $  \tabularnewline
                               & ES2 &a  &$3790.891 $  \tabularnewline
                               &                       &b&$-21.458 $  \tabularnewline
\hline 
dyall.aae3z               &GS              &a  &$4.110$  \tabularnewline
                               &ES1      &a  &$-1.902$  \tabularnewline
                               &                          &b&$14.800 $  \tabularnewline
                               & ES2 &a  &$1348.683 $  \tabularnewline
                               &                       &b&$-19.839 $  \tabularnewline
\hline 
\hline 
\end{tabular}
}
\label{tab:basisset_compare}
\end{table}

\subsection{Enhancement of $E_{\rm PV}$ of H$_2 X _2$}

In this subsection, we numerically verify the enhancement of $E_{\rm PV}$ of \ce{H2$X$2}
in electronic excited states.
For \ce{H2S2}, \ce{H2Se2}, and \ce{H2Te2},
the number of correlated electrons and the size of the active space were limited in our computations.
It is checked how these truncations affected the accuracy by the comparison
between the results of the EOM-CC calculation.
It is also useful to estimate the accuracy of the calculation of larger molecules,
where the active space and the correlated electrons are limited.
The values of $E_{{\rm {PV}}}$ of $\rm{H_2 O_2}$, $\rm{H_2 S_2}$, $\rm{H_2 Se_2}$, and $\rm{H_2 Te_2}$ molecules are summarized in Tables \ref{tab:H2O2_excited}, \ref{tab:H2S2_excited}, \ref{tab:H2Se2_excited}, and \ref{tab:H2Te2_excited}, respectively.
The value of $\lambda$ for \ce{H2$X$2} was chosen
so that the result with the chosen $\lambda$ is well consistent with the value by the Z-vector method. 
The difference of \ce{H2S2} with the $2s2p3s3p$ correlation was about 5\%, 
although that $\lambda = 10^{-3} $ (a.u.) worked well for $3s3p$ correlated calculations.
This small inconsistency is not taken seriously, because the best $\lambda$ for the ground state is not the best one for excited states.

\begin{table*}[p]
\caption{Effect of the active space on $E_{{\rm {PV}}}$ in the $\rm{H_2 O_2}$ molecule. 
	$\lambda=10^{-3}$ (a.u.) was employed. 
}
\centering
\scalebox{1.0}{
	\begin{tabular}{cc.c.....}
		\hline \hline 
		Basis & 
		Correlating~~ & 
		{\rm{Virtual}} & 
		$E_{{\rm {PV}}}/10^{-19}E_{h}$& 
		\multicolumn{1}{c}{} & 
		\multicolumn{2}{c}{$E_{{\rm {PV}}}/10^{-19}E_{h}$} & 
		\multicolumn{2}{c}{}\tabularnewline
		& Orbitals~~ & 
		{\rm{cutoff}}~ & 
		(Z-vector) &  
		& \multicolumn{2}{c}{(FFPT)} &  
		& \tabularnewline
		&  & /E_{h}~~ & 
		GS & 
		\multicolumn{1}{.}{\rm{GS}~} & 
		\multicolumn{2}{.}{\rm{ES1}} & 
		\multicolumn{2}{.}{\rm{ES2}}\tabularnewline
		\cline{5-9} \cline{6-9} \cline{7-9} \cline{8-9} \cline{9-9} 
		&  &  &  &\rm{a}~~&\rm{a}~~&\rm{b}~~~&\rm{a}~~~~&\rm{b}~~~\tabularnewline
		\hline 
		dyall.acv3z & $2s2p$ & $100$ & $4.12$ & $4.04$ & -$1.98$ & $15.39$ & $892.62$ & -$20.94$\tabularnewline
		&  & $500$ & $4.12$ & $4.05$ & -$1.98$ & $15.37$ & $899.28$ & -$20.93$\tabularnewline
		&  & $1500$ & $4.12$ & $4.09$ & -$1.74$ & $15.59$ & $899.43$ & -$20.68$\tabularnewline
		&  & {\rm{All}} & $4.12$ & $4.09$ & -$1.58$ & $15.76$ & $899.69$ & -$20.51$\tabularnewline
		& All & $100$ & $4.14$ & $4.12$ & -$1.79$ & $15.07$ & $1208.00$ & -$19.95$\tabularnewline
		&  & $500$ & $4.15$ & $4.11$ & -$1.88$ & $14.81$ & $1343.10$ & -$19.81$\tabularnewline
		&  & $1500$ & $4.15$ & $4.11$ & -$1.93$ & $14.76$ & $1348.50$ & -$19.85$\tabularnewline
		&  & {\rm{All}} & $4.15$ & $4.08$ & -$2.11$ & $14.59$ & $\bf 1348.72$ & -$20.05$\tabularnewline
		\hline \hline 
	\end{tabular}
}
\label{tab:H2O2_excited}

\caption{Effect of the active space on $E_{{\rm {PV}}}$ in the $\rm{H_2 S_2}$ molecule. 
	$\lambda=10^{-3}$ (a.u.) was employed. 
}
\centering
\scalebox{0.98}{
	\begin{tabular}{cc.c.....}
		\hline \hline 
		Basis & 
		Correlating~~ & 
		{\rm{Virtual}} & 
		$E_{{\rm {PV}}}/10^{-18}E_{h}$& 
		\multicolumn{1}{c}{} & 
		\multicolumn{2}{c}{$E_{{\rm {PV}}}/10^{-18}E_{h}$} & 
		\multicolumn{2}{c}{}\tabularnewline
		& Orbitals~~ & 
		{\rm{cutoff}}~ & 
		(Z-vector) &  
		& \multicolumn{2}{c}{(FFPT)} &  
		& \tabularnewline
		&  &/E_{h}~~& 
		GS & 
		\multicolumn{1}{.}{\rm{GS}~} & 
		\multicolumn{2}{.}{\rm{ES1}} & 
		\multicolumn{2}{.}{\rm{ES2}}\tabularnewline
		\cline{5-9} \cline{6-9} \cline{7-9} \cline{8-9} \cline{9-9} 
		&  &  &  & \rm{a}~~ & \rm{a}~~~ & \rm{b}~~~ & \rm{a}~~~ & \rm{b}~~~\tabularnewline
		\hline 
		dyall.acv3z & $3s3p$ & $10$ & $-1.38$ & -$1.40$ & $354.63$ & $358.98$ & $32.21$ & -$36.67$\tabularnewline
		&  & $50$ & $-1.38$ & -$1.40$ & $355.15$ & $359.46$ & $32.19$ & -$36.71$\tabularnewline
		&  & $100$ & $-1.38$ & -$1.39$ & $355.19$ & $359.49$ & $32.23$ & -$36.66$\tabularnewline
		&  & $1000$ & $-1.38$ & -$1.40$ & $355.17$ & $359.47$ & $32.20$ & -$36.69$\tabularnewline
		&  & $1300$ & $-1.38$ & -$1.40$ & $355.17$ & $359.47$ & $32.20$ & -$36.69$\tabularnewline
		& $2s2p3s3p$ & $10$ &$-1.35$ &-$1.40$ & $365.01$ & $369.10$ & $32.71$ & -$37.23$\tabularnewline
		&            & $50$ & $-1.32$ & -$1.39$ & $368.96$ & $372.53$ & $32.66$ & -$37.83$\tabularnewline
		&            & $100$ & $-1.31$ & -$1.38$ & $368.86$ & $372.41$ & $32.64$ & -$37.83$\tabularnewline
		&            & $1000$ & $-1.31$ & -$1.38$ & $368.94$& $\bf{372.48}$ & $32.65$ & -$37.87$\tabularnewline
		\hline \hline 
	\end{tabular}
}
\label{tab:H2S2_excited}


\caption{Effect of the active space on $E_{{\rm {PV}}}$ in the $\rm{H_2 Se_2}$ molecule. 
	$\lambda=10^{-4}$ (a.u.) was employed. 
}
\centering
\scalebox{0.91}{
	\begin{tabular}{cc.c.....}
		\hline \hline 
		Basis & 
		Correlating~~ & 
		{\rm{Virtual}} & 
		$E_{{\rm {PV}}}/10^{-17}E_{h}$& 
		\multicolumn{1}{c}{} & 
		\multicolumn{2}{c}{$E_{{\rm {PV}}}/10^{-17}E_{h}$} & 
		\multicolumn{2}{c}{}\tabularnewline
		& Orbitals~~ & 
		{\rm{cutoff}}~ & 
		(Z-vector) &  
		& \multicolumn{2}{c}{(FFPT)} &  
		& \tabularnewline
		&  &/E_{h}~~ & 
		GS & 
		\multicolumn{1}{.}{\rm{GS}~} & 
		\multicolumn{2}{.}{\rm{ES1}} & 
		\multicolumn{2}{.}{\rm{ES2}}\tabularnewline
		\cline{5-9} \cline{6-9} \cline{7-9} \cline{8-9} \cline{9-9} 
		&  &  &  & \rm{a}~~ & \rm{a}~~~~ & \rm{b}~~~~ & \rm{a}~~~ & \rm{b}~~~\tabularnewline
		\hline 
		dyall.acv3z & $4s4p$ & $20$ & $-8.08$ & -$8.10$ & $2204.80$ & $2107.95$ & $168.99$ & -$506.46$\tabularnewline
		&  & $30$ & $-8.07$ & -$8.10$ & $2204.86$ & $2108.01$ & $168.98$ & -$506.49$\tabularnewline
		&  & $100$ & $-8.06$ & -$8.10$ & $2204.94$ & $2108.05$ & $168.97$ & -$506.56$\tabularnewline
		&  & $200$ & $-8.06$ & -$8.10$ & $2204.94$ & $2108.05$ & $168.95$ & -$506.57$\tabularnewline
		& $3d4s4p$ & $20$ & $-6.42$ & -$6.49$ & $2281.62$ & $2161.32$ & $167.42$ & -$545.41$\tabularnewline
		&  & $30$ & $-6.24$ & -$6.29$ & $2278.02$ & $2156.20$ & $167.86$ & -$547.23$\tabularnewline
		&  & $100$ & $-6.28$ & -$6.35$ & $\bf{2281.18}$ & $2158.48$ & $167.56$ & -$548.84$\tabularnewline
		\hline \hline 
	\end{tabular}
}
\label{tab:H2Se2_excited}


\caption{Effect of the active space on $E_{{\rm {PV}}}$ in the $\rm{H_2 Te_2}$ molecule. 
	$\lambda=10^{-5}$ (a.u.) was employed. 
}
\centering
\scalebox{0.955}{
	\begin{tabular}{cc.c.....}
		\hline \hline 
		Basis & 
		Correlating~~ & 
		{\rm{Virtual}} & 
		$E_{{\rm {PV}}}/10^{-15}E_{h}$& 
		\multicolumn{1}{c}{} & 
		\multicolumn{2}{c}{$E_{{\rm {PV}}}/10^{-15}E_{h}$} & 
		\multicolumn{2}{c}{}\tabularnewline
		& Orbitals~~ & 
		{\rm{cutoff}}~ & 
		(Z-vector) &  
		& \multicolumn{2}{c}{(FFPT)} &  
		& \tabularnewline
		&  &/E_{h}~~& 
		GS & 
		\multicolumn{1}{.}{\rm{GS}~} & 
		\multicolumn{2}{.}{\rm{ES1}} & 
		\multicolumn{2}{.}{\rm{ES2}}\tabularnewline
		\cline{5-9} \cline{6-9} \cline{7-9} \cline{8-9} \cline{9-9} 
		&  &  &  & \rm{a}~~ & \rm{a}~~~ & \rm{b}~~~ & \rm{a}~~ & \rm{b}~~~\tabularnewline
		\hline
		dyall.acv3z & $5s5p$ & $10$ & $-2.23$ & -$2.23$ & $227.66$ & $235.53$ & $27.68$ & -$103.94$\tabularnewline
		&  & $50$ & $-2.22$ & -$2.22$ & $227.68$ & $235.55$ & $27.73$ & -$103.96$\tabularnewline
		&  & $70$ & $-2.22$ & -$2.23$ & $227.64$ & $235.51$ & $27.68$ & -$104.00$\tabularnewline
		&  & $190$ & $-2.22$ & -$2.23$ & $227.66$ & $235.54$ & $27.71$ & -$103.96$\tabularnewline
		& $4d5s5p$ & $10$ & $-1.87$ & -$1.88$ & $233.61$ & $240.93$ & $30.39$ &-$114.48$\tabularnewline
		&  & $50$ & $-1.85$ & -$1.88$ & $233.97$ & $\bf{241.42}$ & $30.65$ &-$115.37$\tabularnewline
		\hline \hline 
	\end{tabular}
}
\label{tab:H2Te2_excited}
\end{table*}

First, we focus on the enhancement of $E_{\rm PV}$ due to the excitation of the electron. 
In Tables \ref{tab:H2O2_excited}, \ref{tab:H2S2_excited}, \ref{tab:H2Se2_excited}, and \ref{tab:H2Te2_excited},
the most enhanced values with the most reliable computational condition 
are represented in bold-face letters.
The most reliable computational condition was determined based on the effect of the virtual truncation and the correlated electrons,
which are discussed later.
The ratios of the most enhanced values to those in the ground states (Z-vector) with the same computational condition were 324, $-284$, $-363$, and $-130$ for \ce{H2O2}, \ce{H2S2}, \ce{H2Se2}, and \ce{H2Te2}, respectively. 
These results clearly confirm the significant enhancement of $E_{\rm PV}$ due to the electronic excitation.
The enhancement of $E_{\rm PV}$ is much larger than the expected errors discussed in the previous section.
In H$_2$O$_2$, the most enhanced state was ES2, while for H$_2$S$_2$, H$_2$Se$_2$, and H$_2$Te$_2$, the most enhanced states were ES1. One may find $E_{\rm PV}$ of the excited state of a lighter system (e.g., \ce{H2O2}) was larger than that in the ground state of a heavier system (e.g., \ce{H2S2} and \ce{H2Se2}). 
However, this comparison is unfair, 
because the dihedral angle of H$_2$O$_2$ in the optimized geometry was different from other molecules.
It is well known that the values of the PVED are close to zero at $\phi = 90 ^\circ$ \cite{Bast:2011, Senami:2019, Inada:2018, Shee:2016, Thyssen:2000, Laerdahl:1999, Faglioni:2001, Senami:2020, Horny:2015, Stralen:2005}, which is the optimized dihedral angle for H$_2$S$_2$, H$_2$Se$_2$, and H$_2$Te$_2$.
We will discuss this point in detail in the next section.

Next, we discuss the effects of the number of the correlated electrons and the size of the active space on $E_{\rm PV}$ in excited states. The effect of the virtual truncation was less than 1\% except for \ce{H2O2}. In the case of \ce{H2O2}, $E_{\rm PV}$ was sensitive to truncation, while this feature was not found at the ground state. For example, the error of 100 $E_h$ cutoff from the non-cutoff value was about 12\% at ES2(a) when all electrons were correlated. Nevertheless, the truncation of the virtual space did not change the magnitude of the enhancement of $E_{\rm PV}$ significantly. Hence, the use of the small active space is sufficient for the simple estimate of $E_{\rm PV}$, especially for large systems.

For \ce{H2S2}, \ce{H2Se2}, and \ce{H2Te2}
the effect of the correlation of core-valence orbitals were roughly within a few \% for almost all excited states,
and contributions were about 10\% in ES2(b) of \ce{H2Se2} and \ce{H2Te2}.
These contributions were smaller than those in the ground states of \ce{H2Se2} and \ce{H2Te2}, which were about 30\% and 20\%, respectively.
The situation was different in the case of \ce{H2O2}. The contribution of the correlation of the 1s orbital to $E_{{\rm {PV}}}$ was found to be 33.3\% at ES2(a). Hence, the contribution of the core-valence correlation to $E_{\rm PV}$ depends on molecules.
Nevertheless, for ES2(a) of \ce{H2O2}, the enhancement of $E_{\rm PV}$ can be reproduced even in 2s2p correlated computations.
This result encourages us to apply our methodology to large systems, where correlated electrons and the active space are inevitably limited.

\subsection{Mechanism of enhancement of the PVED}

We have confirmed a few hundred times the enhancement of $E_{\rm PV}$ in electronic excited state 
compared to the ground state. 
We should take care of this enhancement,
because the dihedral angles of H$_2$O$_2$ and others are different as 115$^\circ$ and 90$^\circ$, respectively.
Particularly, it is known that the PVED is almost zero around 90$^\circ$.
To clarify how the enhancement occurs, 
we compare the results for the optimized structure with those at $\phi= {45}^\circ$,
where it is one of the most typical targets of the study of the PVED
because of its extremum value of the PVED.

We summarize the values of $E_{\rm PV}$ in the optimized and $\phi = 45^\circ$ structures in Table \ref{tab:H2X2_45_excited}. 
To derive the structure with $\phi = 45^\circ$,
other angles and lengths were fixed on the optimized structure. 
For $\phi = 45^\circ$, most values of $E_{\rm PV}$ in excited states were the same order as the ground state, though the values were increased slightly.
One important observation is that the maximum $E_{\rm PV}$ in excited states of the optimized structure was larger than that in the $ \phi = {45}^\circ$ structure for all \ce{H2$X$2} by one or two orders of magnitude. 
Amazingly, $E_{\rm PV}$ in ES2(a) of \ce{H2O2} in the optimized structure ($1.35\times10^{-16} E_h$)
was larger than that in the ground state of \ce{H2S2} in the $\phi= 45^\circ$ structure ($-1.68\times10^{-17} E_h$),
despite the well-known Z-scaling rule of the PVED, which was first reported in the 1970s \cite{Zel:1977, Harris:1978}. 

\begin{table*}[tb]
	\caption{ $E_{{\rm {PV}}}$ for the optimized and $\phi=45^\circ$ structures of the ${\rm{H_2}}X_2$ molecules. 
	}
	\centering
	\scalebox{1.0}{
		\begin{tabular}{ccccrrrrr}
			\hline \hline
			\multirow{4}{*}{${\rm H_{2}}X_{2}$} & \multirow{4}{*}{Structure} & Correlating & Virtual & \multicolumn{1}{c}{} & \multicolumn{2}{c}{$E_{{\rm {PV}}}/E_{h}$} & \multicolumn{2}{c}{}\tabularnewline
			&  & Orbitals & cutoff &  & \multicolumn{2}{c}{(FFPT)} &  & \tabularnewline
			&  &  & {[}Hartree{]} & \multicolumn{1}{c}{GS} & \multicolumn{2}{c}{ES1} & \multicolumn{2}{c}{ES2}\tabularnewline
			\cline{5-9} \cline{6-9} \cline{7-9} \cline{8-9} \cline{9-9} 
			&  &  &  & a~~~~~~~~~~ & a~~~~~~~~~~ & b~~~~~~~~~~ & a~~~~~~~~~~ & b~~~~~~~~~~\tabularnewline
			\hline 
			${\rm {H_{2}O_{2}}}$ & opt ($\phi=\ang{115}$) & All & All & \,\,$4.08\times10^{-19}$\,\, & \,\,$-2.11\times10^{-19}$\,\, & \,\,$1.46\times10^{-18}$\,\, & \,\,$\bf{1.35\times10^{-16}}$\,\, & \,\,$-2.01\times10^{-18}$\,\,\tabularnewline
			& $\phi=\ang{45}$&  &  & \,\,$-3.45\times10^{-19}$\,\, & \,\,$1.99\times10^{-18}$\,\, & \,\,$\bf{1.74\times10^{-18}}$\,\, & \,\,$1.99\times10^{-18}$\,\, & \,\,$2.73\times10^{-19}$\,\,\tabularnewline
			
			${\rm {H_{2}S_{2}}}$ & opt ($\phi=\ang{90}$) & $2s2p$ & $100$ & \,\,$-1.38\times10^{-18}$\,\, & \,\,$3.69\times10^{-16}$\,\, & \,\,$\bf{3.72\times10^{-16}}$\,\, & \,\,$3.26\times10^{-17}$\,\, & \,\,$-3.78\times10^{-17}$\,\,\tabularnewline
			& $\phi=\ang{45}$&  &  & \,\,$-1.68\times10^{-17}$\,\, & \,\,$1.80\times10^{-17}$\,\, & \,\,$3.55\times10^{-17}$\,\, & \,\,$3.76\times10^{-17}$\,\, & \,\,$\bf{4.01\times10^{-17}}$\,\,\tabularnewline
			
			${\rm {H_{2}Se_{2}}}$ & opt ($\phi=\ang{90}$) & $4s4p$ & $100$ & \,\,$-8.10\times10^{-17}$\,\, & \,\,$\bf{2.20\times10^{-14}}$\,\, & \,\,$2.11\times10^{-14}$\,\, & \,\,$1.69\times10^{-15}$\,\, & \,\,$-5.07\times10^{-15}$\,\,\tabularnewline
			& $\phi=\ang{45}$&  &  & \,\,$-1.63\times10^{-15}$\,\, & \,\,$1.31\times10^{-15}$\,\, & \,\,$\bf{2.13\times10^{-15}}$\,\, & \,\,$2.28\times10^{-15}$\,\, & \,\,$1.64\times10^{-15}$\,\,\tabularnewline
			
			${\rm {H_{2}Te_{2}}}$ & opt ($\phi=\ang{90}$) & $5s5p$ & $70$ & \,\,$-2.23\times10^{-15}$\,\, & \,\,$2.28\times10^{-13}$\,\, & \,\,$\bf{2.36\times10^{-13}}$\,\, & \,\,$2.77\times10^{-14}$\,\, & \,\,$-1.04\times10^{-13}$\,\,\tabularnewline
			& $\phi=\ang{45}$&  &  & \,\,$-2.72\times10^{-14}$\,\, & \,\,$1.23\times10^{-14}$\,\, & \,\,$\bf{3.14\times10^{-14}}$\,\, & \,\,$3.08\times10^{-14}$\,\, & \,\,$1.79\times10^{-14}$\,\,\tabularnewline
			\hline\hline
		\end{tabular}
	}
	\label{tab:H2X2_45_excited}
\end{table*}
\begin{figure}[tbp]
	\centering
	\includegraphics[width=0.93\linewidth]{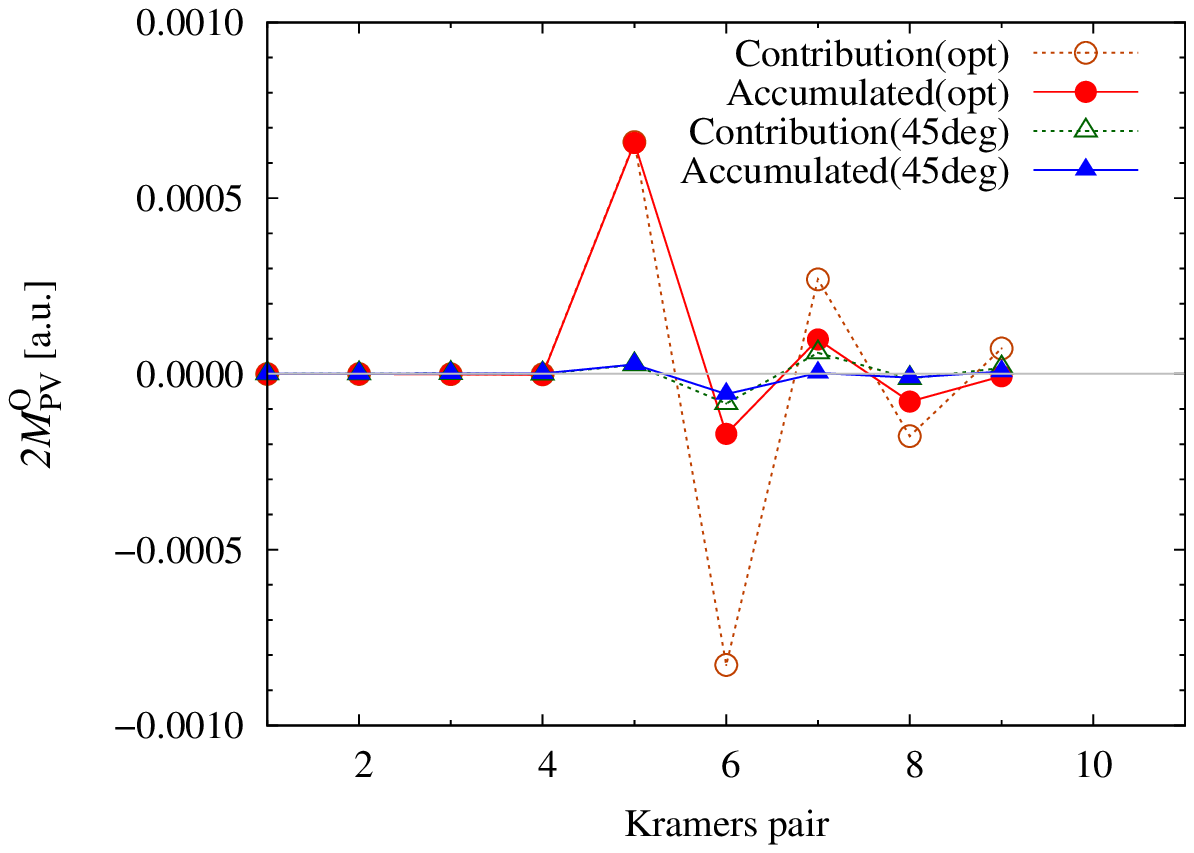} 
	\caption{Contribution to $M_{\rm PV}^{\rm O}$ from each Kramers pair and accumulated value for ${\rm {H_{2}O_{2}}}$
		in the ground state of the optimized and $\phi = 45^\circ$ structures. 
	}
	\label{fig:H2O2_DHF}
\end{figure}
\begin{figure}[tbp]
	\centering
	\includegraphics[width=0.93\linewidth]{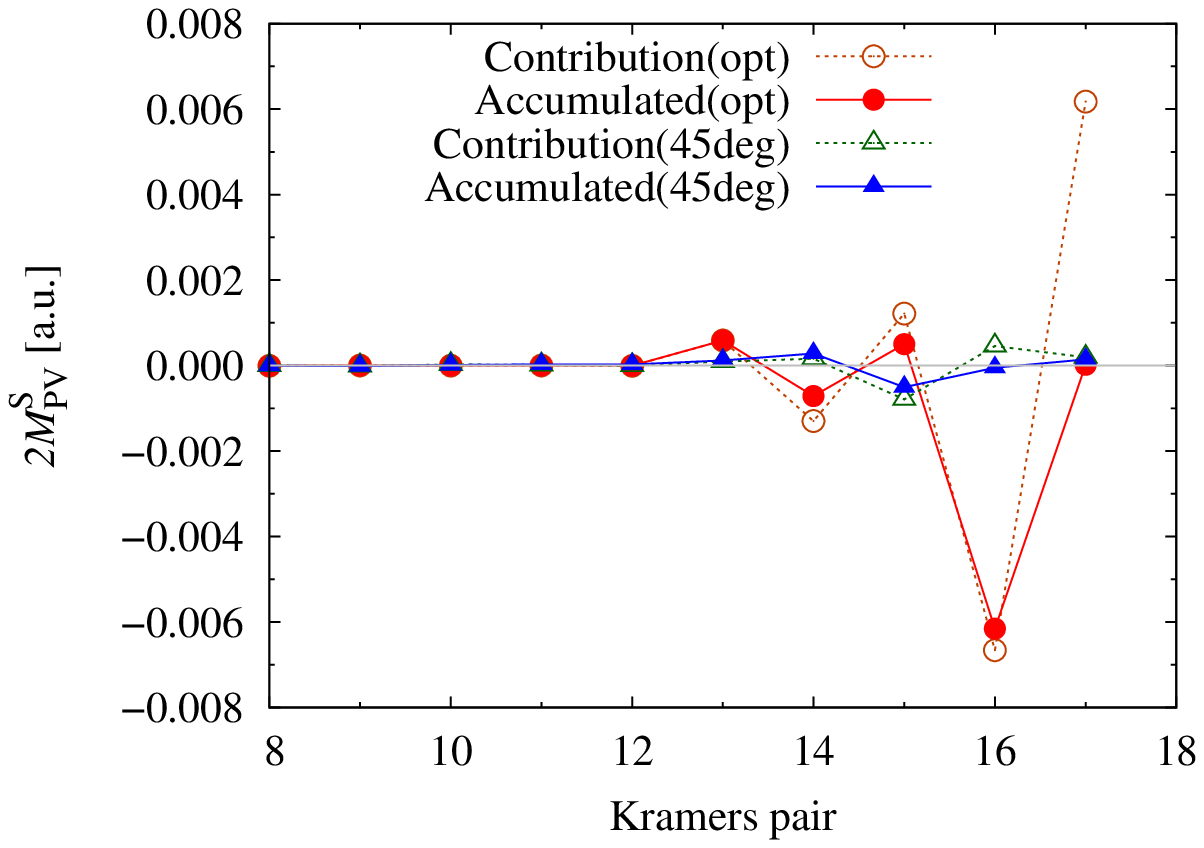} 
	\caption{Contribution to $M_{\rm PV}^{\rm S}$ from each Kramers pair and accumulated value for ${\rm {H_{2}S_{2}}}$
		in the ground state of the optimized and $\phi = 45^\circ$ structures. 
	}
	\label{fig:H2S2_DHF}
\end{figure}
\begin{figure}[tbp]
	\centering
	\includegraphics[width=0.93\linewidth]{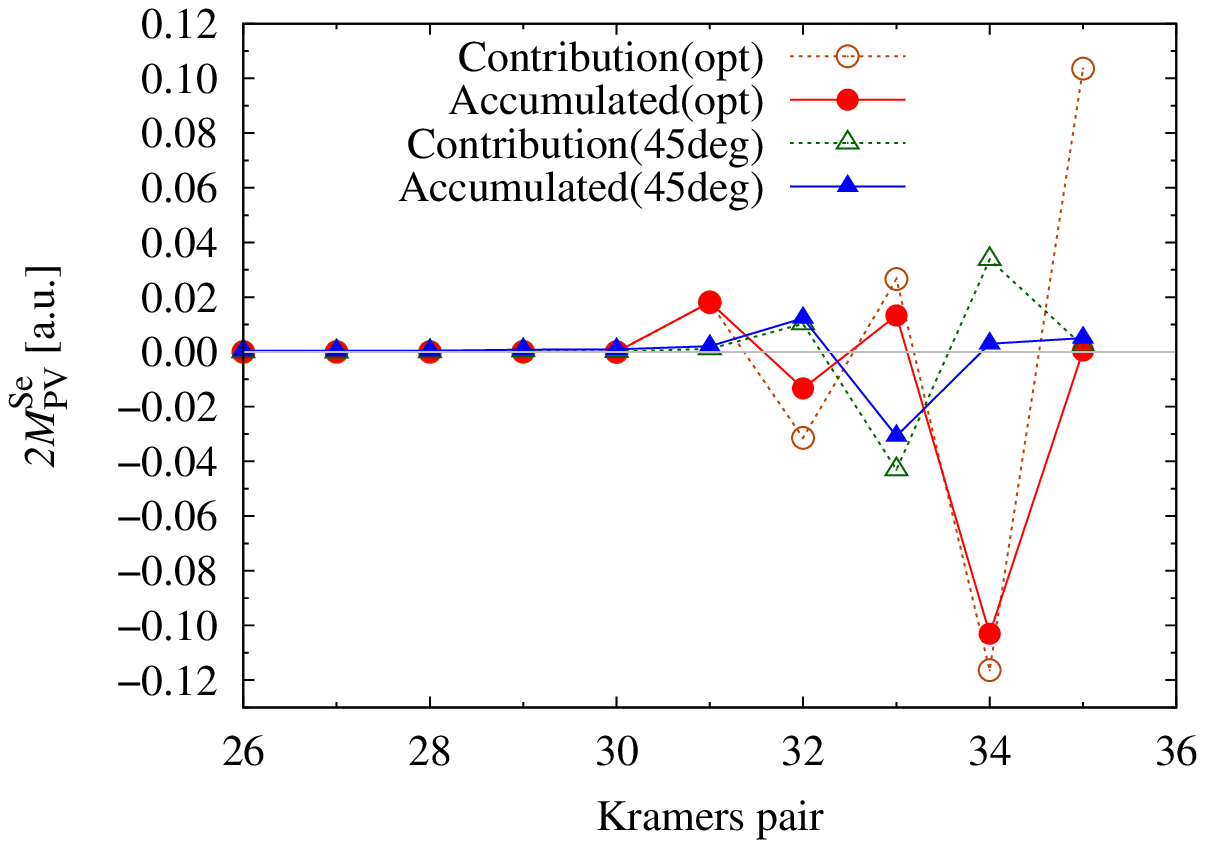} 
	\caption{Contribution to $M_{\rm PV}^{\rm Se}$ from each Kramers pair and accumulated value for ${\rm {H_{2}Se_{2}}}$
		in the ground state of the optimized and $\phi = 45^\circ$ structures. 
	}
	\label{fig:H2Se2_DHF}
\end{figure}
\begin{figure}[tbp]
	\centering
	\includegraphics[width=0.93\linewidth]{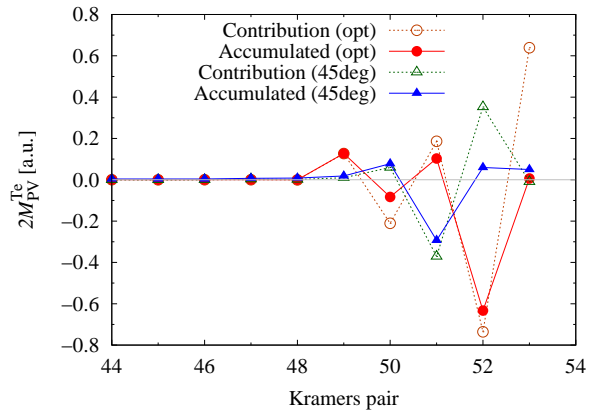} 
	\caption{Contribution to $M_{\rm PV}^{\rm Te}$ from each Kramers pair and accumulated value for ${\rm {H_{2}Te_{2}}}$
		in the ground state of the optimized and $\phi = 45^\circ$ structures. 
	}
	\label{fig:H2Te2_DHF}
\end{figure}

Finally, the enhancement is studied by the contribution, $M^{n,i}_{\rm{PV}}$, from the $i$-th orbital 
to the matrix element of $M^{n}_{\rm{PV}}$. 
The contributions from the $i$-th pair of spinors (Kramers pairs) 2$M^{n,i}_{\rm{PV}}$ and the accumulated total value in the optimized and $\phi={45}^\circ$ structures are shown in Figs. \ref{fig:H2O2_DHF}-\ref{fig:H2Te2_DHF}.
The contributions were calculated by the HF method.
In the figures, the values of 2$M^{n,i}_{\rm{PV}}$ from HOMO to HOMO$-$5 in the optimized structure were larger than those in the $\phi= {45}^\circ$ structure for almost all cases.
HOMO and HOMO$-$1 gave the largest two $M^{n,i}_{\rm{PV}}$ in \ce{H2$X$2} except for \ce{H2O2}, while for \ce{H2O2}, HOMO$-$4 and HOMO$-$5 gave the largest two values.
Despite large $M^{n,i}_{\rm{PV}}$, the total $E_{\rm PV}$ in the ground state of the optimized structure of all H$_2 X_2$ was small because most of them were canceled out each other.

\begin{table*}[t]
	\caption{The total $E_{{\rm {PV}}}$ and the contribution from the HOMO in the ${\rm{H_2}}X_2$ molecules. The ES values of $E_{\rm PV}$ are those in the excited states shown in bold letter in Table \ref{tab:H2X2_45_excited}. 
	}
	\scalebox{1.0}{
		\begin{tabular}{ccrrr}
			\hline \hline
			&  & \multicolumn{3}{c}{ $E_{\rm{PV}}/E_h$ }\tabularnewline
			&  & GS \hspace{7mm} & ES \hspace{7mm} & HOMO \hspace{5mm} \tabularnewline
			\hline
			\ce{H2O2} & opt ($\phi$=\ang{115}) & $4.08 \times10^{-19}$& $1.35 \times10^{-16}$ & $-4.22 \times10^{-18}$\tabularnewline
			& $\phi$=\ang{45} &                           $-3.45 \times10^{-19}$ & $1.74 \times10^{-18}$ & $-9.93 \times10^{-19}$\tabularnewline
			\ce{H2S2} & opt ($\phi$=\ang{90}) & $-1.38 \times10^{-18}$ & $3.72 \times10^{-16}$ & $-7.20 \times10^{-16}$\tabularnewline
			& $\phi$=\ang{45} &                           $-1.68 \times10^{-17}$ & $4.01 \times10^{-17}$ & $-2.25 \times10^{-17}$\tabularnewline
			\ce{H2Se2} & opt ($\phi$=\ang{90}) & $-8.10 \times10^{-17}$ & $2.20 \times10^{-14}$ & $-3.54 \times10^{-14}$\tabularnewline
			& $\phi$=\ang{45} &                             $-1.63 \times10^{-15}$ & $2.13 \times10^{-15}$ & $-7.30 \times10^{-16}$\tabularnewline
			\ce{H2Te2} & opt ($\phi$=\ang{90}) & $-2.23 \times10^{-15}$ & $2.36 \times10^{-13}$ & $-3.73 \times10^{-13}$\tabularnewline
			& $\phi$=\ang{45} &                            $-2.72 \times10^{-14}$ & $3.14 \times10^{-14}$ & $5.59 \times10^{-15}$\tabularnewline
			\hline \hline
		\end{tabular}
	}
	\label{tab:HOMO}
\end{table*}

From these figures, it is speculated that the total $E_{\rm PV}$ increases significantly, if the electron in the orbital with a large $M^{n,i}_{\rm{PV}}$ is excited.
(Here, we simply ignore other contributions, e.g., $M^{n,i}_{\rm{PV}}$ of the virtual orbital to which the electron is excited, and the change of the reference from the HF to the CCSD at the EOM-CC level). 
We analyzed the $R$ vector of the EOM-CC calculation, which describes the excitation from the CC wavefunction, for the most enhanced states in Table \ref{tab:H2X2_45_excited}.
It was found that the excitation from the HOMO accounted for more than 85\%, except for \ce{H2O2} in the optimized structure.
For \ce{H2O2} in the optimized structure, the contribution from the excitation from the HOMO was about 79\%, and that from the excitation from the HOMO$-$1 is about 7\%. 
Table \ref{tab:HOMO} shows the comparison between 
the total $E_{\rm PV}$ for the states in bold letters in Table \ref{tab:H2X2_45_excited}
and the contribution from the HOMO in the ground state. 
For excited states of the optimized geometry,
the HOMO's value was on the order of that of the total $E_{\rm PV}$ of \ce{H2$X$2} except for \ce{H2O2}
and the sign of the PVED is opposite to the contribution from the HOMO.
Accordingly, the magnitude of the maximum enhancement of $E_{\rm PV}$ in the excited state 
may be estimated by $M^{n,i}_{\rm{PV}}$ of the orbital from which the electron is excited.
In Table \ref{tab:HOMO}, for \ce{H2O2},
$E_{\rm PV}$ is much larger than that expected from the HOMO's value.
More detailed analysis may be required and we will study this topic in the future work. 
For the $\phi = 45^\circ $ structure, the values of the HOMO were smaller than or comparable to the total $E_{\rm PV}$ in the ground state.
It would be the reason for the small enhancement in the excited state.

\section{Conclusion}

In this paper, 
the significant enhancement of the PVED ($=2 |E_{\rm PV}|$) by electronic excitation,
which was predicted in Ref.~\cite{Senami:2019},
has been verified for H$_2 X_2$ by accurate computations based on EOM-CC theory. 
The ratio of the most enhanced $E_{\rm PV}$ to that in the ground state 
is 324, $-284$, $-363$, and $-130$ for \ce{H2O2}, \ce{H2S2}, \ce{H2Se2}, and \ce{H2Te2}, respectively. 
It was predicted that electronic excitation or ionization breaks
the cancellation between large contributions to the PVED from orbitals
and then $E_{\rm PV}$ is enhanced \cite{Senami:2019}.
The enhancement has been carefully confirmed by checking 
the dependence on computational parameters and so on.
In the FFPT calculation, 
the suitable value of the perturbation parameter $\lambda$ should be employed for each excited state,
while in this work the value of $\lambda$ is determined by comparing results with those obtained by Z-vector method.
We have confirmed this approach is sufficient for our purpose,
because the dependence on $\lambda$ is much smaller than the enhancement of $E_{\rm PV}$.
In the choice of basis sets,
it has been reported that 
both correlation and diffuse functions are essential to 
computations of $E_{\rm PV}$ in excited states.
Diffuse functions are important for ordinary excited state computations
and crucial for the computation of the PVED.
$E_{\rm PV}$ in ES2(a) with the dyall.ae3z basis set was calculated 
three times as large as that with dyall.aae3z.
The PVED in excited states was unaffected by 
the truncation of active space and the correlation of core-valence orbitals
except for \ce{H2O2}.
The difference between different truncations was within 1\%.
The deviation by different correlation of core-valence orbitals
was within 3\% for almost all excited states,
while this deviation was about 30\% and 20\% in the ground states of \ce{H2Se2} and \ce{H2Te2}, respectively.
In \ce{H2O2}, the difference by the truncation was about 12\% in the largest case,
and the contribution of the correlation of the 1s orbital was about 33.3\% in ES2(a). 
Nevertheless, 
the enhancement of $E_{\rm PV}$ was much larger than expected errors discussed in this paper. 
Although our results of H$_2$O$_2$ were sensitive to the choice of active space and correlation orbitals,
it was almost insensitive for other H$_2X_2$.
From the above consideration, 
it is encouraged to apply our methodology to large systems,
such as chiral molecules with heavy elements and amino acid molecules.
From the study of the contribution to the PVED from each orbital,
it has been speculated that 
$E_{\rm PV}$ is remarkably enhanced 
if electrons are excited from the orbital with a large contribution
which is canceled out with contributions from other orbitals.
In our results,
electrons were excited from the HOMO dominantly.
The contribution to $E_{\rm PV}$ from the HOMO in the ground state of the optimized structure 
had a similar size to the total $E_{\rm PV}$ in the most enhanced excited states except for H$_2$O$_2$.
Hence, we have proposed the hypothesis that 
the maximum enhancement of $E_{\rm PV}$ by electronic excitation
can be estimated by the contribution to $E_{\rm PV}$ from the orbital from which the electron is excited.
This speculation was true for \ce{H2$X$2} except \ce{H2O2},
while the enhancement in \ce{H2O2} was much larger than the expected value from our hypothesis.
Hence the enhancement mechanism in H$_2$O$_2$ requires further investigation,
and we will study the enhancement mechanism of the PVED in detail in our future work.

The significant enhancement found in this work is speculated to occur for other chiral molecules.
The enhancement of the PVED by the electronic excitation originates in 
the breaking of the cancellation between large contributions from orbitals.
This cancellation in the ground state is the generic property for most chiral molecules
as discussed in Ref.~\cite{Bast:2011}.
Therefore, this enhancement may be the key to discovering 
the imprint of the weak interaction in chiral molecules experimentally.


\begin{acknowledgments}
This work was supported by Grants-in-Aids for Scientific Research (17K04982, 19H05103, and 21H00072).
A.S. acknowledges financial support from the Japan Society for the Promotion of Science (JSPS) KAKENHI Grant No. 20K22553 and 21K14643.
We are also thankful to the supercomputer of ACCMS (Kyoto University) for the main calculation
and Research Institute for Information Technology, Kyushu University (General Projects).
\end{acknowledgments}

\end{document}